\documentclass[lettersize,journal]{IEEEtran}
\usepackage{amsmath,amsfonts}
\usepackage{algorithmic}
\usepackage{algorithm}
\usepackage{array}

\usepackage{textcomp}
\usepackage{stfloats}
\usepackage{url}
\usepackage{verbatim}
\usepackage{graphicx}
\usepackage{cite}
\usepackage{booktabs}  
\usepackage{arydshln}  
\usepackage{siunitx}
\usepackage{graphicx}
\usepackage{subcaption} 
\usepackage{caption} 
\usepackage{siunitx} 
\usepackage{float} 
\usepackage{graphicx}

\begin{document}

\title{Compact Varactor-Integrated RIS for Wideband and Continuously Tunable Beamforming}


\author{Yikun Li,~\IEEEmembership{Member,~IEEE,}
        Yufei Zhao,~\IEEEmembership{Member,~IEEE,} 
        Chau Yuen,~\IEEEmembership{Fellow,~IEEE,}
        Xiong Qin,
        Christ Clenson,
        Chao Du,~\IEEEmembership{Member,~IEEE,} 
        and~Yong Liang Guan,~\IEEEmembership{Senior Member,~IEEE}
\thanks{Manuscript received, Jan., 2025; revised **, ** 2025. (Corresponding author: Yufei Zhao, e-amil: yufei.zhao@ntu.edu.sg).}
\thanks{Yikun Li, Yufei Zhao, Chau Yuen, Xiong Qin, and Yong Liang Guan are with the School of Electrical and Electronic Engineering, Nanyang Technological University, 639798, Singapore.

Christ Clenson is with the Department of Electronic and Telecommunication Engineering, University of Moratuwa, 10400, Sri Lanka.

Chao Du is with the Multifunctional Materials and Structures, Key Laboratory of the Ministry of Education \& International Center for Dielectric Research, School of Electronic Science and Engineering, Xi’an Jiaotong University, Xi’an, 710049, China.
}}

\markboth{Journal of \LaTeX\ Class Files,~Vol.~*, No.~*, Mar.~2025}%
{Shell \MakeLowercase{\textit{et al.}}: A Sample Article Using IEEEtran.cls for IEEE Journals}

\IEEEpubid{0000--0000/00\$00.00~\copyright~2021 IEEE}

\maketitle

\begin{abstract}
This letter presents a novel Reconfigurable Intelligent Surface (RIS) that features a low-profile structure, wide operating bandwidth, and continuous phase control. By incorporating a middle patch layer without introducing an additional air gap, the proposed design maintains a thin form factor, while achieving a smooth 310° phase shift over 10\% bandwidth at 6.1 GHz with excellent reflection. A fabricated 10×10 RIS array exhibits stable performance, enabling precise beam control across a 600 MHz bandwidth. These results highlight the potential of the proposed low-profile, wideband RIS with continuous phase tuning for next-generation wireless communication systems.

\end{abstract}


\section{Introduction}
\IEEEPARstart
{R}{econfigurable} Intelligent Surfaces (RIS) have emerged as a transformative technology in wireless communications, enabling dynamic wavefront manipulation with programmable electromagnetic (EM) properties \cite{ref1,ref2,ref3,ref4}. However, practical deployment poses challenges, particularly the need for structurally low-profile designs that maintain stable performance across diverse environments, including mobile platforms such as vehicles and UAVs \cite{ref5,ref6,ref7}. Ensuring high-efficiency reflection \cite{ref8} and wideband operation \cite{ref9} under these conditions is crucial for reliable wireless communications, driving the needs of a RIS design with low profile, wideband, and continous phase control for precise control. 

Existing RIS designs face fundamental challenges in achieving three key properties essential for advanced wireless communication: low-profile structure, wideband operation, and continuous phase-shifting capability. Many RIS architectures attempt to address these aspects, but trade-offs often limit their ability to excel in all three simultaneously. The first challenge lies in maintaining a low-profile structure while integrating tunable components. Conventional approaches often require additional biasing circuits or multilayer configurations, increasing structural thickness \cite{ref10} and complicating large-scale deployment, particularly in space-constrained applications such as vehicular communication systems \cite{ref11}. To improve bandwidth performance, researchers have explored the use of wideband materials \cite{ref12} and tunable amplifiers \cite{ref13}. However, these approaches frequently introduce additional circuit complexity \cite{ref14} and higher power consumption, which can be detrimental to low-profile RIS designs. Furthermore, integrating such active components typically requires additional bias circuit, further limiting scalability and increasing design overhead. Achieving continuous phase control presents further challenges, as existing technologies come with inherent trade-offs. Various phase-tuning methods, such as liquid crystals (LCs) \cite{ref15}, vanadium dioxide (VO$_2$) \cite{ref16}, and micro-electro-mechanical systems (MEMS)-based RIS \cite{ref17}, have been explored. While LCs and VO$_2$ provide continuous phase modulation, they suffer from slow response times \cite{ref18} and limited phase tuning ranges \cite{ref19}, making them unsuitable for dynamic, high-speed applications. MEMS-based designs enable precise and continuous phase tuning but are fundamentally optimized for operation at a single frequency, leading to relatively narrow bandwidth \cite{ref20}. As a result, achieving all three properties—low-profile integration, wideband operation, and seamless phase control—remains a major challenge in RIS development, limiting practical deployment. To address this, an RIS that simultaneously ensures low-profile, broadband, and continuous phase control is crucial in space-constrained environments. Such a design enables seamless integration while maintaining high reflection efficiency and broad phase-tuning capability. It also provides a practical solution for applications requiring precise real-time reconfigurability. By balancing these three properties, the proposed RIS enhances wireless communication performance, making it a strong candidate for future mobile and networking applications.

In this letter, we propose a varactor-based RIS with a low-profile design, featuring a total thickness of less than \( \lambda_0 / 10 \), where \( \lambda_0 \) is the wavelength at the central operating frequency of \SI{6.1}{\giga\hertz}. The low-profile structure ensures seamless integration into space-constrained environments. The proposed RIS also achieves a broad 10\% operating bandwidth, while maintaining the amplitude response exceeding 70\%, making it suitable for robust wireless applications. To further enhance performance, we introduce a novel middle patch in our design, which enables the following key properties:

\begin{itemize}
    \item A low-profile structure without additional air gap.
    \item A wideband seamless response of both phase and amplitude across the operating range.
    \item A continuous tunability over the working band.
    
\end{itemize}

\newpage

Our proposed varactor-based RIS integrates these three properties in a single RIS architecture represents a significant advancement, offering a practical solution for real-world applications in vehicular communications, satellite networks, and high-mobility scenarios.

\begin{table}[h]
    \centering
    \small 
    \renewcommand{\arraystretch}{1.3} 
    \setlength{\tabcolsep}{3pt} 
    \caption{Comparison of previous works and this work.} 
    \label{tab:comparison}
    \resizebox{\columnwidth}{!}{ 
    \begin{tabular}{>{\centering\arraybackslash}m{3.5cm}  
                    >{\centering\arraybackslash}m{1cm}  
                    >{\centering\arraybackslash}m{1cm}  
                    >{\centering\arraybackslash}m{1cm}  
                    >{\centering\arraybackslash}m{1cm}  
                    >{\centering\arraybackslash}m{1cm}  
                    >{\centering\arraybackslash}m{1.3cm}} 
        \toprule
        \textbf{Property} & \textbf{[8]} & \textbf{[10]} & \textbf{[15]} & \textbf{[16]} & \textbf{[19]} & \textbf{This work} \\
        \midrule
        \textbf{Low-profile} \\ (thickness \(< \lambda_0/10\)) 
        & \raisebox{2.5ex}{\checkmark} &  & \raisebox{2.5ex}{\checkmark} & \raisebox{2.5ex}{\checkmark} &  & \raisebox{2.5ex}{\checkmark} \\
        \hdashline
         \vspace{2mm} \textbf{Wideband} \\ (bandwidth \( \geq \) 10\%) \vspace{1mm}
        &  &  &  & \raisebox{2.5ex}{\checkmark} & \raisebox{2.5ex}{\checkmark} & \raisebox{2.5ex}{\checkmark} \\
        \hdashline
        \vspace{2.5mm} \textbf{Continuous phase} \\ \textbf{control} \vspace{1mm}
        &  & \raisebox{2.5ex}{\checkmark} & \raisebox{2.5ex}{\checkmark} &  &  & \raisebox{2.5ex}{\checkmark} \\

        \bottomrule
    \end{tabular}
    }
\end{table}

\section{RIS Design and Simulation}

The RIS unit cell model and dimensions are first proposed, followed by an analysis of design principles for reducing structural thickness, expanding the operating bandwidth, and improving continuous phase tunability. The simulation results of amplitude and phase, along with the schematic of the control board, are presented sequentially.

\subsection{Unit Cell Design and Simulation}

\begin{figure}[t]
    \centering
    \vspace{0mm}

    \subfloat[]{%
        \includegraphics[width=1\linewidth]{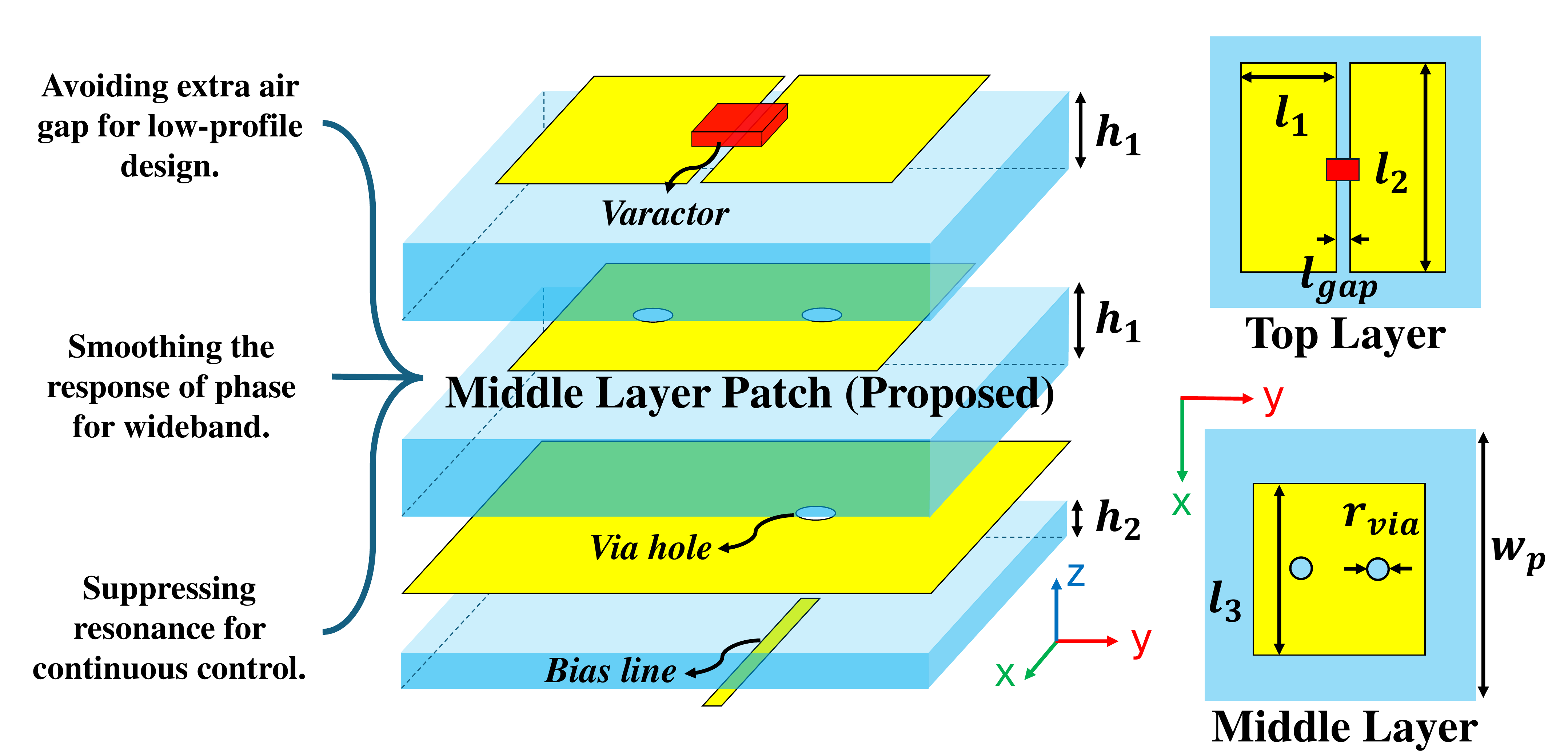}
        \label{Schematic Model}
    }\\ 

    \vspace{1mm} 

    \subfloat[]{%
        \raisebox{-0.7mm}{\includegraphics[width=0.485\linewidth]{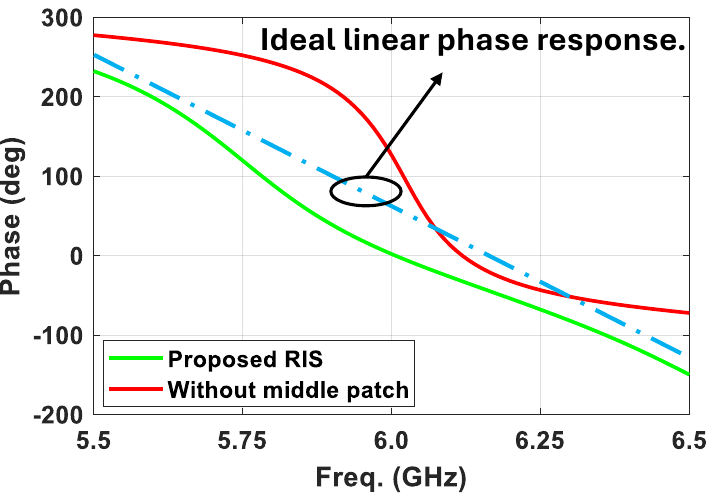}} 
        \label{Phase with and without}
    }
    \subfloat[]{%
        \includegraphics[width=0.475\linewidth]{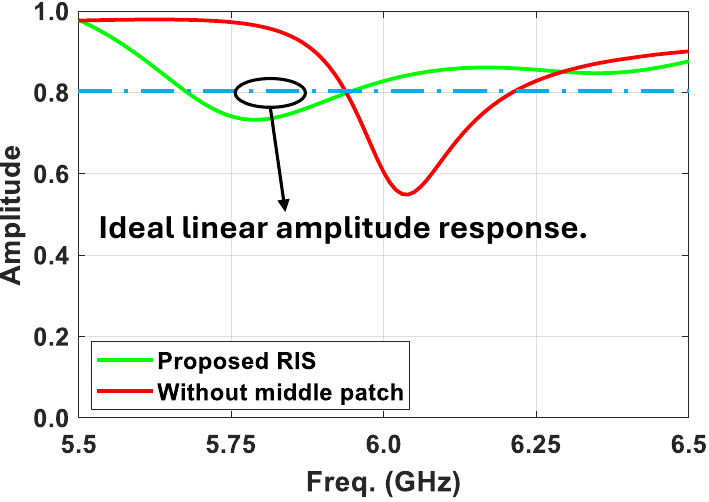}
        \label{Amplitude with and without}
    }

    \caption{
         (a) 3D and 2D model schematic. The detailed dimensions are $h_{\textrm{1}}=1.524\,\text{mm}$, $h_{\textrm{2}}=0.813\,\text{mm}$, $l_{\textrm{1}}=4.2\,\text{mm}$, $l_{\textrm{2}}=10\,\text{mm}$, $l_{\textrm{3}}=8.5\,\text{mm}$, $l_{\textrm{gap}}=0.2\,\text{mm}$, $r_{\textrm{via}}=0.4\,\text{mm}$, $w_{\textrm{p}}=13.5\,\text{mm}$, and a total thickness of $3.9\,\text{mm}$. (b) Phase smoothing achieved with the middle patch. (c) Amplitude enhancement using middle patch.
    }
    \label{fig_2}
\end{figure}

Each unit cell is designed with a square geometry, measuring \SI{13.5} \times \SI{13.5}{\milli\meter^2} or $0.27\times0.27\,\lambda_0^2$ at the central operating frequency of \SI{6.0}{\giga\hertz}. To simplify the configuration of bias lines controlling the varactor, they are positioned beneath the third layer, which serves as the ground. This arrangement effectively shields the influence of the bias lines, minimizing electromagnetic interference and ensuring stable performance. All substrate layers of the proposed unit cell are composed of Rogers 4003 ($\epsilon_r=3.55$, $\tan\delta=0.0027$), a low-loss dielectric material well-suited for wireless communications. Two holes are created for placing metal vias: one connects the varactor on the top layer to the bias lines on the bottom layer, while the other connects the varactor to the ground, ensuring reliable electrical continuity. Additionally, a radio frequency (RF) inductor is integrated into each bias line to prevent RF signal degradation and ensure that the unit cell's RF performance remains uncompromised by the DC feedings. The schematic model of the unit cell is illustrated in Fig.\!~\ref{fig_2}(a).

\subsubsection{Low-Profile Design Without Additional Air Gap}
Conventional wideband RIS designs often require an additional air gap to enhance bandwidth, which increases the overall thickness of the structure. In contrast, the proposed approach eliminates this necessity by introducing a compact and simple second patch layer, which achieves bandwidth expansion, while preserving the low-profile nature of the unit cell. This ensures a total thickness of less than \( \lambda_0 / 10 \), making it ideal for applications where seamless integration are essential. By avoiding extra spacing layers, the proposed structure maintains a thin and efficient form factor.

\captionsetup{font=small}

\subsubsection{Wideband Operation via Smooth Resonance Tuning}
The integration of the middle patch layer enables a wider operating bandwidth by leveraging resonance damping. In the equivalent circuit model, a patch is primarily represented as a capacitance. Introducing the patch increases the overall capacitance of the unit cell, leading to a dampened resonance that suppresses abrupt variations in reflection phase drop. The resonance condition follows \cite{ref21}:
\begin{equation}
f_{\textrm{res}}=\frac{1}{2\pi\sqrt{LC}},\quad \textrm{BW}\propto\frac{C}{L}.
\end{equation}
By increasing the unit cell capacitance, the proposed structure achieves a more stable reflection response across the frequency band, mitigating resonance-induced distortions. This results in an expanded bandwidth as demonstrated in Fig.~\ref{fig_2}(b). The improved resonance behavior ensures consistent performance across the working band, making the phase response more linear against frequency variations.

\subsubsection{Continuous Phase Control for Precise Beamforming}
\begin{figure}[h] 
    \centering
    \vspace{0mm}
    \hspace{0mm} 
    \subfloat[]{%
        \includegraphics[width=0.495\linewidth]{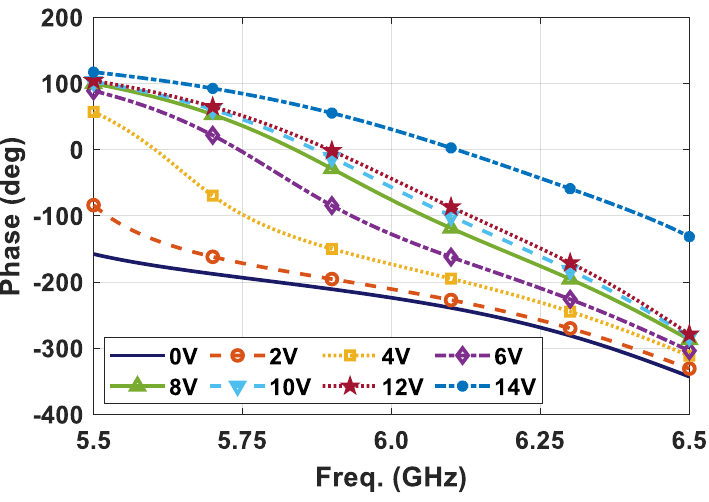}%
        \label{fig:image1}
    }
    \subfloat[]{%
        \includegraphics[width=0.48\linewidth]{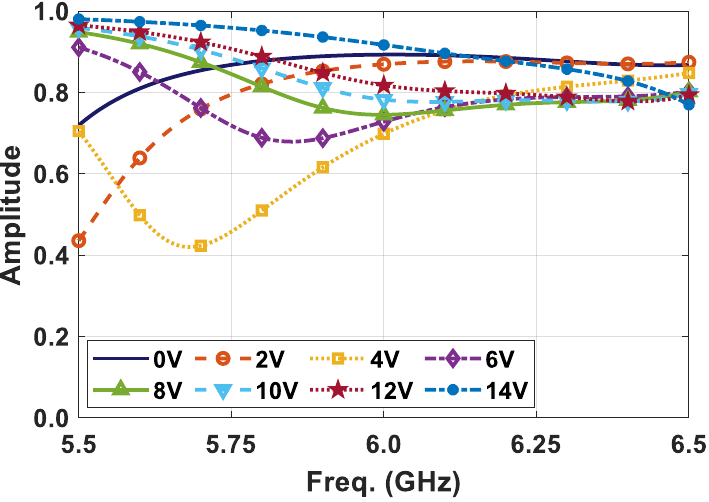}%
        \label{fig:image2}
    }
    \vspace{0mm} 
    \caption{Simulation performance of the proposed RIS unit cell in ANSYS HFSS. (a) Amplitude. (b) Phase.}
    \label{fig_3}
    \vspace{-2mm} 
\end{figure}

The integration of the middle layer patch enhances bandwidth, while also ensuring a more stable amplitude distortion. In the proposed RIS structure, continuous phase tuning is achieved using varactor diodes strategically placed between the patches in the top layer. The Skyworks SMV1231-079LF varactor diode, selected for its stable capacitance-voltage characteristics and low-loss properties, enables reliable phase modulation. A well-designed control board independently adjusts the reflection response of each unit cell. The SPICE-modeled equivalent circuit in \cite{ref22} ensures accurate simulation, with the controlled voltage varying from \SI{0}{V} to \SI{14}{V}. As shown in Fig.~\ref{fig_2}(c), varactor-based tuning enables relatively linear amplitude response. Fig.~\ref{fig_3}(a) and (b) further illustrate the unit cell’s phase and amplitude performance under different bias voltage, demonstrating smooth resonance behavior that minimizes abrupt phase drops, which shows precise phase tuning ensures adaptability to dynamic wireless environments, facilitating high-accuracy beam steering for next-generation reconfigurable intelligent surfaces, with a maximum phase difference of \ang{310}.

\subsection{Control Board Design}
\begin{figure}[hb]
    \centering
    \hspace{-4.5 mm} 
    \includegraphics[width=0.8\linewidth]{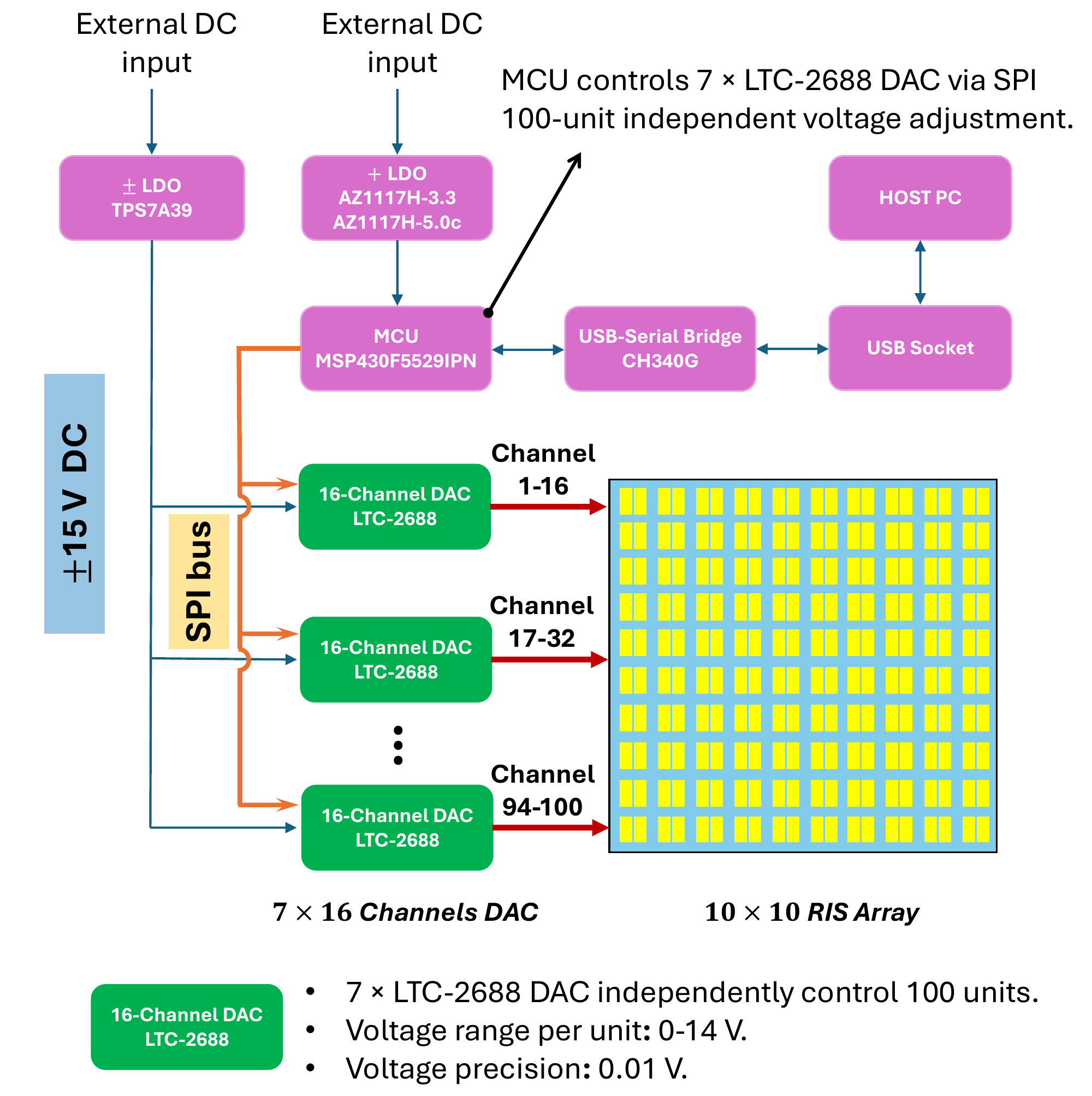} 
    \caption{Circuit schematic of control board.}
    \label{fig_control}
    \vspace{-4mm} 
\end{figure}

The control board was designed to precisely and independently regulate the voltage of 100 varactor-based unit cells. To achieve this, seven LTC-2688 Digital-to-Analog Converters (DACs) were deployed, each offering 16 programmable channels and a \SI{0}-\SI{14}{\volt} output range with \SI{0.01}{\volt} precision. The MSP430F5529 microcontroller acts as the central controller, efficiently managing all seven DACs to ensure simultaneous, independent voltage control for each unit cell. The communication framework leverages UART (MATLAB to MSP430F5529) for high-level control and SPI (MSP430F5529 to LTC-2688) for seamless data transmission, ensuring real-time and high-precision voltage adjustments. This modular and scalable design enhances system flexibility, making it well-suited for dynamic and reconfigurable applications. The detailed schematic of the control board is depicted in Fig.~\ref{fig_control}.

\section{Fabrication and Measurement}
\subsection{Fabrication of prototype}
\begin{figure}[h]
    \centering
    \vspace{0mm}

    \subfloat[Control board.]{%
        \includegraphics[width=1\linewidth]{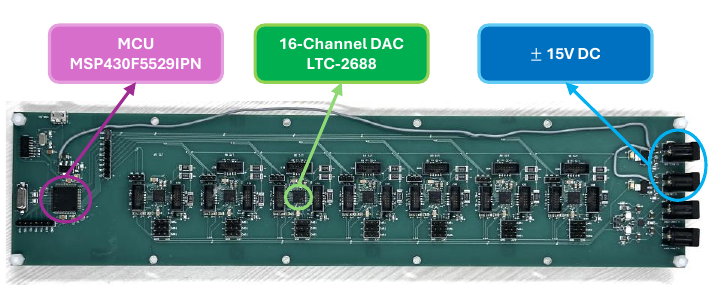}
        \label{fig:control_board}
    }\par 

    \vspace{4mm} 

    \subfloat[Side view of proposed RIS.]{%
        \includegraphics[width=1\linewidth]{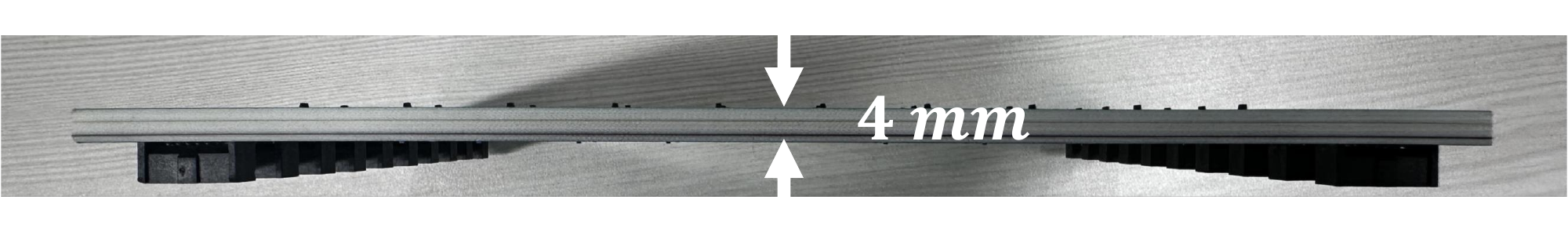}
        \label{fig:top_view}
    }\par 

    \vspace{4mm} 

    \subfloat[Top view of proposed RIS.]{%
        \includegraphics[width=0.46\linewidth]{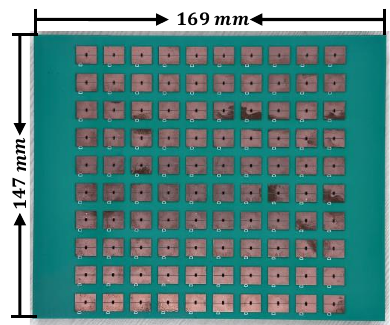}
        \label{fig:bottom_view}
    }
    \
    \subfloat[Bottom view of proposed RIS.]{%
        \includegraphics[width=0.46\linewidth]{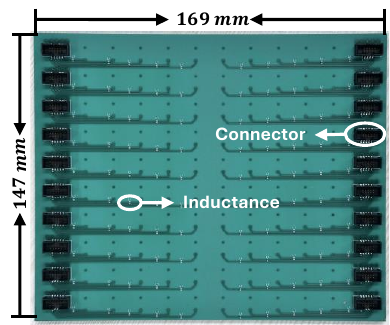}
        \label{fig:another_view}
    }

    \caption{Fabricated prototypes of the control board and RIS board.}
    \label{fig_5}
\end{figure}

The fabricated prototype of the control circuit board, top and bottom view of proposed RIS board is depicted in Fig.~\ref{fig_5}(a), (b), and (c). With additional 20 connectors at the edge of the RIS board, the total size is \SI{169} \times \SI{147} \times \SI{4} {\milli\meter^3} or $3.38\times2.94\times0.08\,\lambda_0^3$. Voltage adjustments from \SI{0}{\volt} to \SI{14}{\volt} are facilitated by LTC-2688, with each controlling 16 unit cells independently, resulting in seven DACs arranged on the control board. The control board features 10 input ports, a ground port, and a USB connector for computer command inputs via Python, allowing precise control over voltage provided to each unit cell.

\captionsetup{font=small}

\subsection{Setup and measurement}
\begin{figure}[h]
    \centering
    \hspace{-2.5 mm} 
    \includegraphics[width=0.98\linewidth]{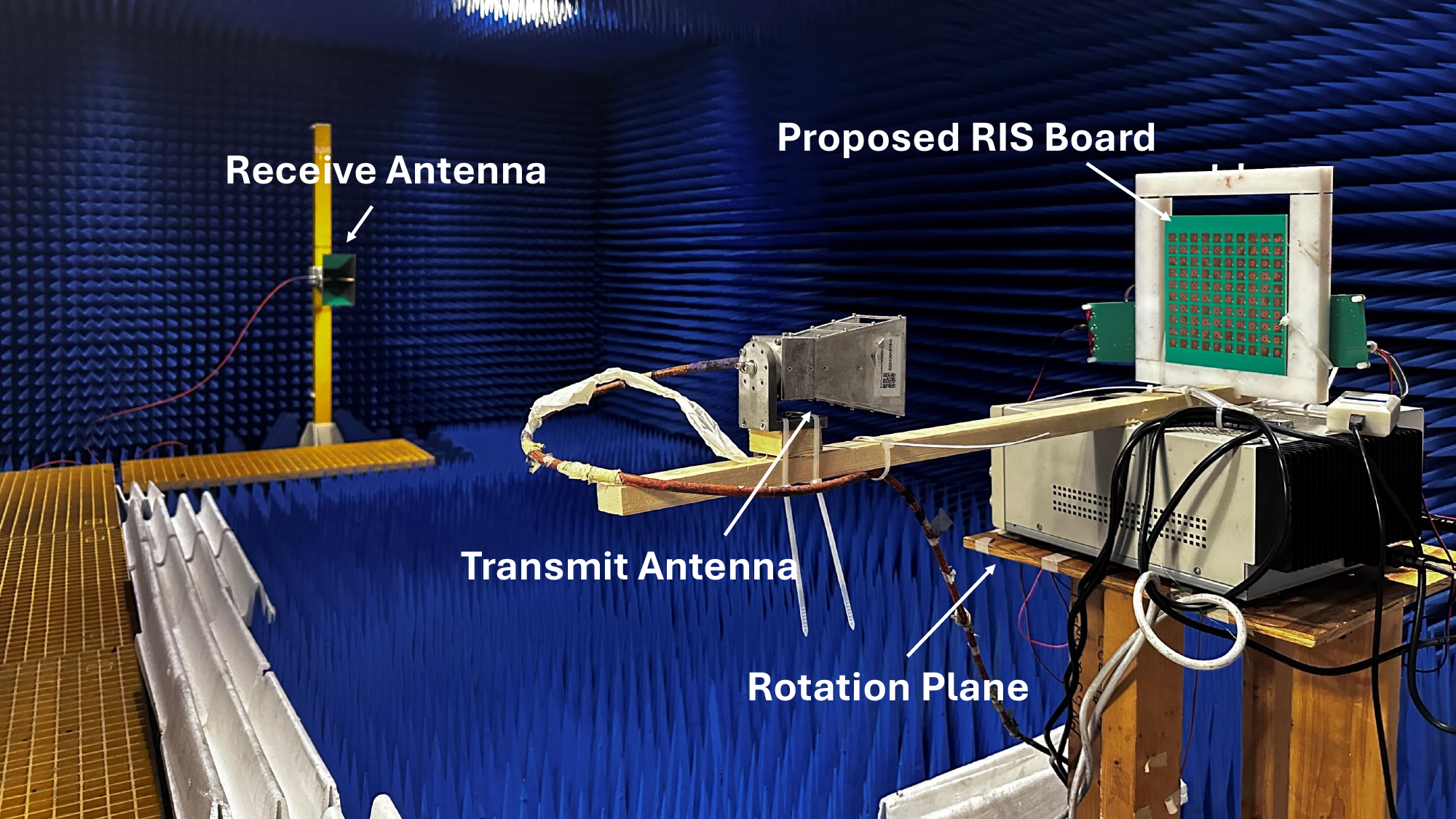} 
    \caption{Measurement setup.}
    \label{fig_7}
    \vspace{-4mm} 
\end{figure}

\
 The measurement setup for the RIS system is depicted in Fig.~\ref{fig_7}. A pair of standard-gain horn antennas serve as the transmitting (Tx) and receiving (Rx) units, both connected to a Vector Network Analyzer (VNA) for precise characterization of the reflection and scattering properties. The RIS dynamically steers a vertically oblique incident wave by adjusting the array distribution according to precomputed phase profiles. During measurements, the Tx antenna remains fixed at a predetermined position, while the RIS reflection phase is modified in real time via a digitally controlled bias network. For near-field evaluations, the Tx horn is placed \SI{450}{\milli\meter} from the RIS center and rotated from \ang{-90} to \ang{90} with a step of \ang{1}, allowing for accurate beam characterization across different incidence angles. To minimize environmental interference and undesired reflections, radio-frequency (RF) absorbing materials are strategically placed around both the Tx and Rx antennas, as well as on the surrounding test fixtures. The RIS system, comprising the designed array, connectors, and RLC components, is securely held within a 3D-printed support structure, ensuring stable positioning and repeatability in measurements. Additionally, the control circuit board is positioned adjacent to the RIS setup to facilitate precise voltage regulation, real-time phase tuning, and seamless integration with the measurement system. To ensure high measurement accuracy, all cables and connectors are carefully calibrated, and background noise is mitigated through a series of baseline measurements \cite{AWPL1,AWPL2,IOT}.

\subsection{Results and Discussions}

\begin{figure}[h] 
    \centering
    
    \vspace{2mm} 

    \captionsetup{font=footnotesize} 
    \subfloat[\footnotesize{Beam towards \ang{15} at \SI{5.8}{\giga\hertz}.}]{%
        \includegraphics[width=0.48\linewidth]{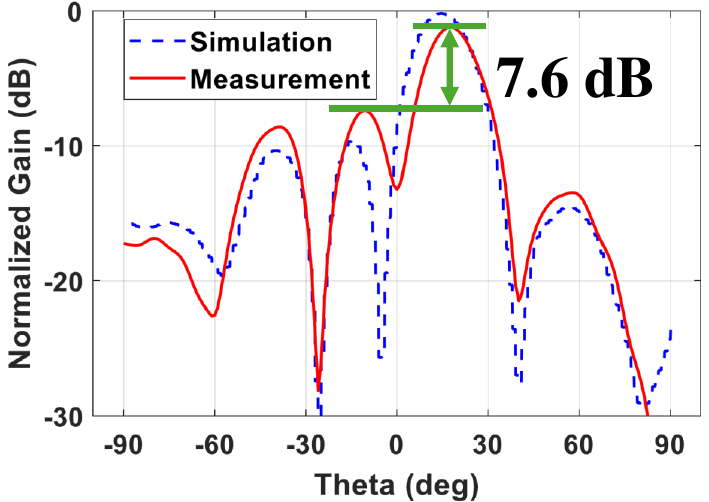}%
    }
    \hfill
    \subfloat[\footnotesize Beam towards \ang{30} at \SI{5.8}{\giga\hertz}.]{%
        \includegraphics[width=0.48\linewidth]{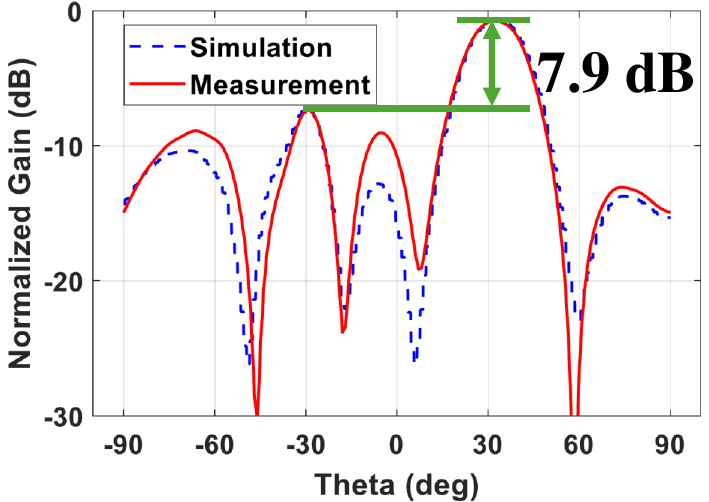}%
    }

    \vspace{2mm} 
    \subfloat[Beam towards \ang{15} at \SI{6.1}{\giga\hertz}.]{%
        \includegraphics[width=0.48\linewidth]{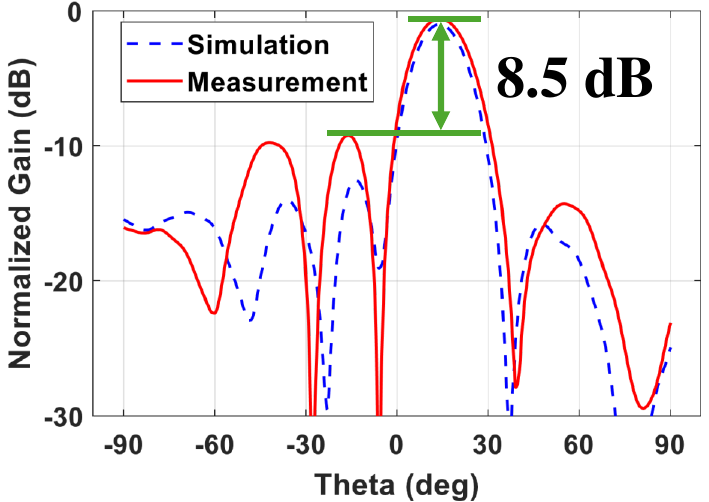}%
        \label{fig:image3}
    }
    \hfill
    \subfloat[Beam towards \ang{30} at \SI{6.1}{\giga\hertz}.]{%
        \includegraphics[width=0.48\linewidth]{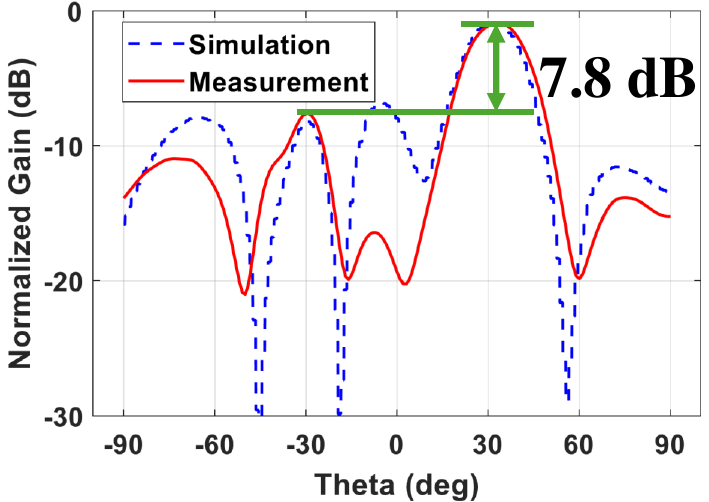}%
        \label{fig:image4}
    }

    \vspace{2mm} 
    \subfloat[Beam towards \ang{15} at \SI{6.4}{\giga\hertz}.]{%
        \includegraphics[width=0.48\linewidth]{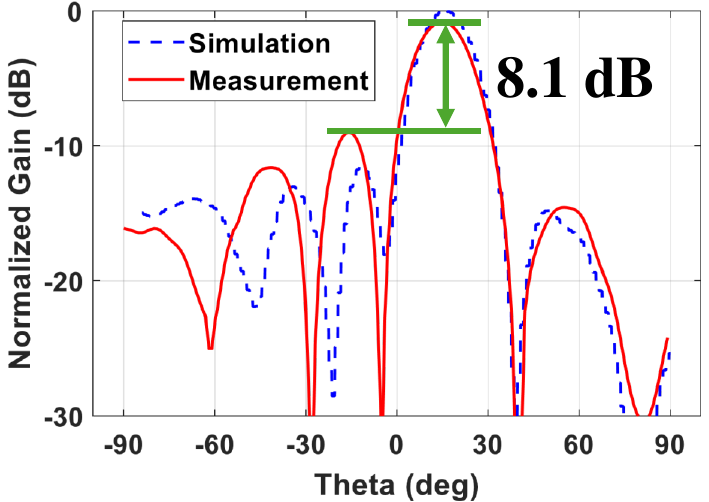}%
        \label{fig:image5}
    }
    \hfill
    \subfloat[Beam towards \ang{30} at \SI{6.4}{\giga\hertz}.]{%
        \includegraphics[width=0.48\linewidth]{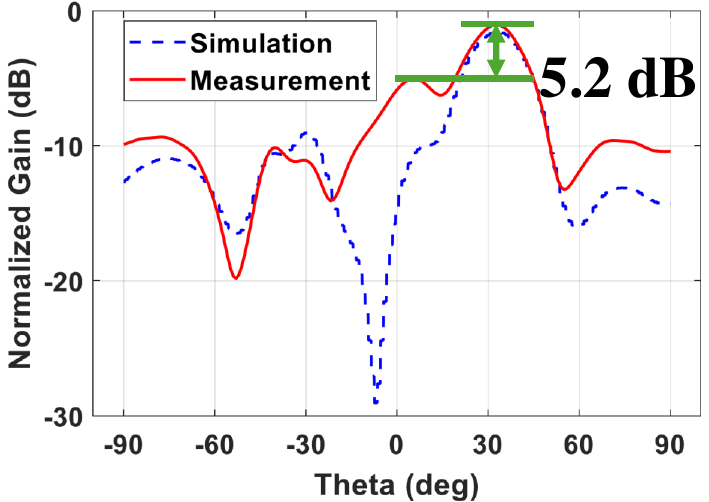}%
        \label{fig:image6}
    }

    \caption{Radiation patterns of directional beam across \SI{5.8} to \SI{6.4}{\giga\hertz}.}
    \label{fig_4}

    \vspace{-1mm} 
\end{figure}

The phase distribution across the array was independently computed and applied along the polarized direction based on the equation in \cite{ref23}. Directional reflection beams were designed for frequencies ranging from \SI{5.8} to \SI{6.4}{\giga\hertz}, targeting deflection angles of \ang{15} and \ang{30}, as illustrated in Fig.~\ref{fig_4}. The simulated radiation patterns reveal that the side-lobe levels are approximately \SI{8}{\decibel} lower than the main lobes, ensuring high beam-steering accuracy with minimal radiation loss. Moreover, the directivity and normalized gain remain highly consistent across the operating bandwidth, confirming stable beamforming performance across different frequencies. The compact size allows the proposed RIS design to be efficiently integrated into space-constrained environments, while effectively achieving precise beam-steering, maintaining high directivity, and robustly suppressing side-lobe levels. Additionally, the ability to maintain stable performance across a broad bandwidth enhances its suitability for multi-frequency and multi-user wireless networks, where dynamic beam adjustment is crucial for optimizing signal quality and minimizing interference.

While simulation and measurement results generally align, slight discrepancies are observed in side-lobe levels and minor beam shifts. These variations likely arise from fabrication tolerances, misalignment in the measurement setup, and parasitic effects within the control circuit. Further refinement of the bias network, improved shielding techniques, and enhanced calibration procedures could help mitigate these effects and improve measurement accuracy, ensuring closer agreement between theoretical and experimental results.

\section{Conclusion}
In this letter, we introduce a Reconfigurable Intelligent Surface (RIS) with low-profile structure, wideband operation, and continuous phase control. The proposed RIS works at \SI{6.1}{\giga\hertz} and achieves a near \ang{310} phase shift, while maintaining a high reflection amplitude of $70\%$. A $10 \times 10$ RIS array was designed, simulated, and experimentally validated for directional beam steering. Total thickness of our RIS is less than \( \lambda_0 / 10 \), ensuring seamless deployment in space-constrained environments.

The fabricated prototype successfully generated reflection beams at \ang{15} and \ang{30}, with measured main lobes approximately \SI{8}{\dB} latger than the side lobes, demonstrating effective beam steering capability. The experimental results closely align with simulations, validating the proposed RIS's ability to achieve precise phase control and stable reflection performance. Our work confirms the effectiveness of the design for next-generation wireless communication systems.


%

\vfill

\end{document}